\documentclass[prl,aps,12pt,wideabs]{revtex4}
\usepackage{amsmath,amssymb,graphicx,mathrsfs,hyperref,slashed}
\newcommand{\be}{\begin{equation}}
\newcommand{\ee}{\end{equation}}
\newcommand{\bea}{\begin{eqnarray}}
\newcommand{\eea}{\end{eqnarray}}

\def\({\left(} \def\){\right)}

\begin{document}

\title{
{Quantum gravitational effects suppress the formation of trapped surfaces}}
\author{\large Ram Brustein${}^{(1)}$,  A.J.M. Medved${}^{(2,3)}$,
Hagar Meir${}^{(1)}$
\\
\vbox{
\begin{flushleft}
 $^{\textrm{\normalsize
(1)\ Department of Physics, Ben-Gurion University,
   Beer-Sheva 84105, Israel}}$
$^{\textrm{\normalsize (2)\ Department of Physics \& Electronics, Rhodes University,
 Grahamstown 6140, South Africa}}$
$^{\textrm{\normalsize (3)\
The National Institute for Theoretical and Computational Sciences  (NITheCS), 
South Africa}}$
\\ \small \hspace{0.57in}
   ramyb@bgu.ac.il,\  j.medved@ru.ac.za, hagarmei@post.bgu.ac.il
\end{flushleft}
}}
\date{}


\begin{abstract}

Classical general relativity predicts, as established by Christodoulou,  that a contracting, spherically symmetric matter system with a large-enough mass will result in the formation of a trapped region whose outer boundary is an apparent horizon where the gravitational redshift diverges. The incompleteness theorems of Penrose and Hawking then lead to the conclusion that the outcome of the collapse is the singular geometry of a Schwarzschild black hole. Both analyses rely on solving Einstein's equations, which constitute a set of partial differential equations that are valid in the limit that the Schwarzschild radius is finite but the Planck length is set to zero, so that quantum fluctuations of the geometry are completely absent. Here, we keep both parameters finite, allowing the geometry to fluctuate quantum mechanically, and take the limit of vanishing Planck length only at the end.
Expressing the geometry of a spherically symmetric, collapsing, thin shell of matter in terms of an effective quantum field theory in 1+1 dimensions, we show -- using the standard techniques of quantum field theory in curved spacetime -- that the quantum width of the would-be horizon scales as the shell's Schwarzschild radius even after enforcing the limit of vanishing Planck length. This finite width implies that the quantum expectation value of the product of the scalar expansions
for the associated null vectors is never vanishing. The conclusion is that an apparent horizon does not form even when the shell has reached its gravitational radius. As the collapse continues, the classical Schwarzschild geometry can no longer be used to describe the shell's exterior geometry.  This provides the sought-after loophole that is needed to explain how astrophysical black holes could be compact objects that are regular and horizonless.

\end{abstract}
\maketitle


\section{Introduction}

Paradoxical issues like information loss \cite{info}, causality violation \cite{mathurCP} and ``firewalls'' \cite{AMPS} are  part and parcel to the paradigm of semiclassical black holes (BHs). The semiclassical approximation in this context means that the Planck length $\;l_P=\sqrt{\hbar G}\;$ is set to zero, eliminating completely the quantum fluctuations of the geometry, while the Schwarzschild radius $\;R_S=2MG\;$ is kept finite, $M$ being the  BH mass.  In this approximation, one includes, in addition to  the classical BH solutions of general relativity,  quantum fields in the curved background \cite{BD} and Hawking radiation \cite{Haw}.

A way around these conundrums would be if astrophysical BHs were actually described by ultracompact but regular objects \cite{carded}. Such BH mimickers \cite{thumper} would also have to  be free of an apparent horizon, which is a localized  surface of infinite redshift, and be able to reproduce most, if not all, of the established properties of a BH in a state of equilibrium. A key challenge that any model of BH mimickers needs to confront is  the seemingly inevitable formation of a singularity when a sufficiently massive system is subject to gravitational collapse, as established by  the celebrated  theorems of Penrose and Hawking \cite{PenHawk1,PenHawk2,HawkEl} and the variations thereof \cite{sing}.

To circumvent these theorems requires that one of their requisite assumptions is not valid for collapsing matter.   A key assumption is the existence of a closed trapped surface; that is, a spacelike two-surface whose orthogonal null geodesics both have negative scalar expansions \cite{HawkingEllis}.  The apparent horizon is the outer boundary of the trapped region and is a null surface for which at least one of the two expansions must be vanishing and, therefore, a surface of infinite redshift. This assumption is  difficult to evade given that a trapped surface does itself appear to be a universal feature of gravitational collapse and could form over arbitrarily large length scales. This was first shown long ago in a model-specific way  by Oppenheimer and Snyder \cite{OS} (also see \cite{MTW}). Decades later, Christodoulou, in an influential series of articles \cite{Chris1,Chris2,Chris3,Chris4,Chris5,Chris6} (also, \cite{Chris7}),  has firmly established the inevitability of a trapped surface dynamically forming during the collapse of matter with regular initial conditions.

Christodoulou's results have since been substantiated by numerical methods \cite{numeric1,numeric2,numrev} and a (very incomplete) list of extensions  include \cite{dust1,dust2,aniso1,aniso2,charge1,charge2,adSV,charge3,higgsV,refine1,refine2,refine3,dSV}.
For a comprehensive review on the overall program, see \cite{review}.

The analysis leading to the proofs of the singularity and trapped-surface formation theorems  rely on solving the Einstein equations, which constitute a set of partial differential equations (PDEs) whose validity necessitates that  the Planck length is set to zero.
Whereas, in our analysis, we keep the Planck length finite, allowing the geometry to fluctuate quantum mechanically, with the limit of vanishing Planck length being imposed  only at the end. Our main objective is to calculate a scalar quantity that can be regarded as a quantum ``horizon-order parameter'': the expectation value of the product of the two scalar expansions.~\footnote{ The concept of horizon-order parameter was first introduced in \cite{hop}.} By incorporating quantum fluctuations, we can take into account the effect of the excitation of the metric by gravitational particle production  on the values of the expansions as the would-be-horizon is approached. This particle production can be attributed to the  motion of the collapsing shell.

Previously, the backreaction of the produced particles was discussed and shown to be too weak to affect the motion of the shell in a significant way ({\em e.g.}, \cite{paddy,manbear}), concluding that the formation of a trapped surface  is inevitable. This conclusion relied on the implicit assumption that, as the shell continues to collapse, a trapped surface does form. Our results in \cite{shellgame} support the estimates of the backreaction. However, by calculating the influence of the produced particles, which is dominated by those produced near the would-be horizon,  directly on the quantum expectation value of the product of  the two expansions, we show that it remains finite in the limit of vanishing Planck length.

Our conclusion is that the classical, PDE-based framework for establishing the formation of a trapped surface breaks down as the matter shell approaches the would-be horizon.  The reason for the breakdown is not because the  backreaction of the produced particles is so strong but, rather, the significant change to our geometric indicator ---  the  horizon-order parameter  --- informing us that a trapped surface has not formed. The purpose of the current work is to explain and support this conclusion

From this perspective, the singularity theorems can no longer be viewed as an impediment to proposals for modelling collapse from regular initial conditions on standard-model matter, such as baryons, electrons and photons, to regular BH mimickers.

Units with $\;R_S=2MG=1\;$ are often employed, with $M$ being  the mass of the matter shell.
A longer companion paper \cite{shellgame} contains additional results and
a much more detailed account of the analysis.

\section{Effective 2D field theory for a contracting shell}

Following  the same route as Christodoulou and many others, we consider four-dimensional (4D) Einstein gravity as sourced by a  spherically symmetric shell of matter of negligible thickness contracting along a null trajectory. The metric inside the shell is that of Minkowski space and the one outside is that of Schwarzschild. As the shell contracts, the radial position of the  boundary between the two geometries is thus changing with time.

We promote $r$ in the angular elements  of our spherically symmetric metric from its status as a radial coordinate to that of a spacetime scalar field $\Phi$, the dilaton,
\begin{equation}
    ds^2 = \gamma_{\alpha\beta}dx^\alpha dx^\beta +\Phi^2(x_\alpha,\theta,\phi)d\Omega ^2\;,
\end{equation}
where $\; \Phi = \sum_{\ell, m} \phi^\ell _m (U,V) Y^m _\ell (\theta,\phi)\;$
corresponds to the 4D scalar field whose square is describing the area of spheres at a given value of the radius $r$.
We assume that only $\;\Phi_0\equiv\phi_0\;$ is classically nonvanishing.~\footnote{The inclusion of some low-$\ell$ modes in addition to the zero mode is not expected to significantly alter the results.}

The action contains the   Einstein--Hilbert Lagrangian along  with the Lagrangian for the matter shell,
\begin{equation}
    S_4 \;=\; \frac{1}{16\pi l_P^2 }\int d^2x d\Omega^2 \sqrt{-g}(R +\mathcal{L}_{matter})\;.
\end{equation}
By integrating over the angular coordinates, we dimensionally reduce this 4D theory of Einstein gravity to a 2D  dilaton--gravity theory.  The dimensional reduction also introduces a large number of degenerate scalar fields $\phi^\ell_m(x_\alpha)$ whose physical origin  is the 4D angular modes perturbing the spherical surface. In the 2D theory, they are scalars, whose labels $\ell$, $m$ are counting labels, not angular quantum numbers.

To proceed, we reexpress $\Phi^2$ as $\Phi\Phi^*$
and regard the $\phi^\ell _m$ fields as real. The result is
\bea
    S_2 \;&=&\; \frac{1}{l_P^2} \int d^2x \sqrt{-\gamma}  \sum\limits_{\ell,m}  \biggl\{ \frac{1}{2} +\frac{1}{4} R^{(2)} (\phi^\ell_m)^2  \cr &+& \frac{1}{2}\gamma^{\alpha\beta} \partial_\alpha\phi^\ell_m \partial_\beta \phi^\ell_m +\frac{1}{4} (\phi^\ell_m)^2 \mathcal{L}_{matter}\biggr\}\;.
    \label{action1}
\eea
Importantly, angular derivatives of $\Phi$ are absent in this action, as these would have integrated to yield  the constant Euler characteristic of the two-sphere plus a total derivative whose integral on the sphere vanishes \cite{reduction1,reduction2,reduction3,mycft,carlip}. This absence of angular derivatives is specific to the dimensional reduction of the Einstein action.

The  equation of motion for each of the fields $\phi^\ell_m$ can be derived from the action (\ref{action1}),
\begin{equation}
    \Box \phi_\ell -\frac{1}{2}R^{(2)}\phi_\ell -\frac{1}{2}\phi_\ell \mathcal{L}_{matter} \;=\; 0\;.
    \label{eomx}
\end{equation}
To simplify the notation, we have denoted $\phi^\ell_m$ by $\phi_\ell$.

The mass squared of $\phi_\ell$ is given by,
\be
m^2(\phi_{\ell})\;=\;\frac{\sqrt{-\gamma}}{2}\left(\frac{1}{2}R^{(2)} +\frac{1}{2} \mathcal{L}_{matter}\right)\;.
\ee

Each of the modes inherits a time-dependent mass  which
arises because of the contraction of the shell,
\be
\frac{1}{2}R^{(2)}(r)=
\begin{cases}
			0\;,\;\;\; & t < t_{shell}\;,\\
            \frac{2MG}{r^3}\;,\;\;\; & t> t_{shell}\;,
\end{cases}
\label{m2t1}
\ee
where $t_{shell}$ means the time $t$ for which $\;r_{shell}(t)=r\;$.
In what follows, we ignore the contribution of the time-independent mass. This mass encodes the interaction of the produced particles with the background  geometry because  $\frac{2MG}{r^3}$ is the Schwarzschild gravitational potential for a scalar field.
Ignoring the effects of this mass means that we neglect the backscattering of the particles from the geometry. We also checked that adding its contribution does not make a significant change to our results.

In terms of the tortoise radial coordinate
$\;r^*=r+\ln(r-1)\;$ along with the Schwarzschild
time $t$ (which is asymptotically Minkowski time),
the equation of motion becomes that of a canonically normalized field in a 2D flat spacetime,
\be
\left(-\partial_t^2+\partial_{r^*}^2 -m^2(r^*)\right) \phi_\ell\;=\;0\;,
\ee
with a revised mass-squared function of the form~\footnote{The PL function is defined in the Appendix.}
\begin{equation}
    m^2(r^*) \;=\;  \dfrac{PL(e^{r^*-1})}{(1+PL(e^{r^* -1}))^4} \;=\;\frac{r-1}{r^4}\;.
    \label{m2Rstar}
\end{equation}

In the tortoise case, we may expand the $\phi_{\ell}$'s in a basis of modes which look like those of a flat-space basis,
\be
\phi_\ell(k) \;=\; \phi_k(t) e^{i k r^*}\;,
\label{flatmode}
\ee
where we have omitted the $\ell$ label on the right  to avoid clutter.
Consequently, $\phi_k(t)$ solves the standard equation
\begin{equation}
\left(-\partial_t^2 -k^2 -m^2(r^*)\right) \phi_k(t)\;=\; 0\;.
\label{dispersion}
\end{equation}
The field must satisfy the asymptotic boundary condition,
\be
\phi^{in}_{k}(t)\;=\; \tfrac{1}{\sqrt{2 \omega }} e^{-i \omega t}\;, \ \ t\;\to\; -\infty\;,
\ee
with  $t$ being the asymptotic Minkowski time.
For the ingoing null trajectory,  $\;dt + dr^*=0\;$, so that  $\;m^2(r^*)=m^2(-t)\;.$~\footnote{The last expression assumes a suitable choice of integration constant.}

\section{Particle production by a contracting thin shell}

We now consider the production of $\phi_{\ell>0}$ particles.  The motion of the shell excites the $\phi_\ell$ fields from their ground state to some excited state which can be characterized by its non-vanishing occupation number.

To make contact with the established methods \cite{BD,offord}, it is helpful if the $\phi_\ell$'s are recast in their canonically normalized form $\;\phi_{\ell}\to \widetilde{\phi}_{\ell} = {\frac{1}{l_P}}\phi_{\ell}\;$.We will use the dimensionless modes $\widetilde{\phi}_{\ell}$ but denote these as $\phi_{\ell}$ or just $\phi$ to avoid clutter.

Our interest is in calculating the variance or two-point function $\langle \phi_{\ell}^2 \rangle$,~\footnote{These are equivalent, $\;\Delta \phi_{\ell}^2 = \langle \phi_{\ell}^2  \rangle\;$, as the expectation value of any of these modes must be vanishing by virtue of the spherically symmetric background.}
\bea
&& \langle\phi_{\ell}^2\rangle (t)\;=\;\tfrac{1}{2\pi}\int dk \frac{1}{|k|} |\beta_k(t)|^2 \;.
\label{varoft1}
\eea
An approximate expression for $\beta_k$ is \cite{offord}
\bea
\beta_k &=& \frac{i}{2 k} \int_{t_{i}}^{t_{{f}}} dt \, m^2(t) e^{-2ikt}
      \; = \;\frac{i}{2 k} \int_{r_{\min}}^{r_i} dr \, \frac{1}{r^3} e^{-2ikt(r)} \;, \nonumber \\
\eea
where a subscript of $i$ denotes initial, one of $f$ means final and
$r_{min}$ determines the minimal shell radius. For a null trajectory, $\;t(r)=-r^*=-\left(r+\ln(r-1)\right)\;$ so that
$\;\frac{dt}{dr}=\frac{dt}{\;dr^*}\frac{\;dr^*}{dr}=-\frac{\;dr^*}{dr}=-\frac{r}{r-1}\;$.

The variance  of a single original mode  with $\;\ell>0\;$ is order $l^2_P$, as expected
for perturbatively small corrections to the zero mode $\;\langle\Phi^2_0\rangle = r^2\;$.
To estimate the total variance, we need to multiply by the number of different $\phi_{\ell}$ fields, $\;{\cal N}_{\ell}= \sum_{\ell =1}^{\ell_{max}}\sum_{m=-\ell}^{\ell}1
\simeq \ell_{max}^2 \;. $
As for  $\ell_{max}$,  this should be the inverse of the smallest discernible angular variation on the surface of the shell, which is $\sqrt{4\pi r^2}/l_{uv}$ for a shell of radius $r$  such that $l_{uv}$ is the ultraviolet cutoff scale for the theory. As the theory is 4D Einstein gravity, the cutoff scale is $l_P$. It follows that $\;{\cal N}_{\ell}(r) = \frac{r^2}{l_P^2}\;$, where the $4\pi$ has not been included as we have already
integrated over the angles to arrive
at Eq.~({\ref{action1}).   Thus, the result for the total variance then scales as $r^2$, so that the width of the would-be horizon scales as  $R_S$, the horizon radius itself!

Turning to the explicit calculation for a single mode
\bea
&& \langle\phi_{\ell}^2\rangle (t)=
\int\limits_{t_i}^t dt_1~m^2(t_1)\int\limits_{t_i}^t dt_2~m^2(t_2)
\\ &&\times
\tfrac{1}{2\pi}\int\limits^{\infty}_{-\infty}~ dk~\frac{1}{4|k|^3} e^{-2 i k (t_1-t_2)~} \;,\nonumber
\label{varoft}
\eea
and
introducing an  infrared cutoff $k_{min}$, we find that, for the right-most integral,
\bea
\int\limits_{0}^\infty\! \frac{dk}{4\pi}~ \frac{\cos\left[2 k|t_1-t_2|\right]}{(k^2+k_{min}^2)^{3/2}}
\;=\;
\frac{|t_1\!-\!t_2|^2~ K_1(2 k_{min}|t_1\!-\!t_2|)}{4\pi~ k_{min}|t_1\!-\!t_2|}\;,\ \ \
\label{kvar}
\eea
where $K_1$ is a modified Bessel function for which  $\;\frac{1}{4\pi} \frac{K_1(2 k_{min}|t_1-t_2|)}{k_{min}|t_1-t_2|}\simeq \frac{1}{3}\;$.~\footnote{We use an approximation
of $\;k_{min}|t_1-t_2|=\frac{1}{\pi}\;$ that is motivated in \cite{shellgame}.}
This leads to an expression for the total quantum variance,
\bea &&
 \langle\Phi_q^2(r_{min})\rangle_{4D}= \frac{r^2_{min}}{3} \\ &\times&\int\limits_{r_{min}}^{r_{i}} \int\limits_{r_{min}}^{r_{i}}
dr_1~dr_2~\frac{1}{r_1^3}~ \frac{1}{r_2^3}~
\left|r_2-r_1+\ln[\frac{r_2-1}{r_1-1}]\right|^2\;. \nonumber
\eea

The integration of the above yields a complicated and not very illuminating expression. However,  comparing the magnitude of the contribution of the three terms in the absolute value squared,
one finds  that the dominant contribution is from the first term $(r_2-r_1)^2$ whose integral for $\;r_i \gg r_{min}\;$ is given by $\frac{1}{r_{min}^2} \ln[\frac{r_i}{r_{min}}]$. Using this approximation and restoring the units, we obtain
\bea
\langle\Phi_q^2(r_{min})\rangle_{4D}
\;=\;  \frac{1}{3} (2MG)^2  \ln[\frac{r_i}{r_{min}}]\;.
\label{varHHH}
\eea

As for the potentially divergent logarithm,  we have in \cite{shellgame} adopted a complementary analysis, which  incorporates the time-independent part of the mass to all orders, to obtain an upper bound  on $\langle\Phi_q^2\rangle_{4D}$
that is finite for  $\;k_{min}>0\;$ and  independent of $r_i$.
We conclude that a non-perturbative treatment would swap the logarithm for an
order-unity number.

In any event,  the variance remains quite small even for  very large values of $r_i$.  Taking, as an example, the collapse of a solar-sized star with $\;r_i/r_{min}\simeq 2\times 10^5\;$, the ratio $\langle\Phi_q^2\rangle/\langle\Phi_0^2\rangle$ becomes order unity only when the shell is close to the would-be horizon, as well it should.~\footnote{For the purpose of this estimate, it is important to include the subleading terms.}

\section{Expansion parameters at the would-be horizon}

Classically, the product of the  two expansion parameters $\Theta_U \Theta_V$ vanishes at the horizon because $\;\Theta_V=0\;$ while $\Theta_U$ is finite.
Outside of the horizon, this product is negative due to the positivity of
$\Theta_V$ and the negativity of $\Theta_U$. See the Appendix for their formal definitions,
along with some pertinent analysis.

As discussed in the Introduction, the product $\Theta_U \Theta_V$ can be viewed as a type of
horizon-order parameter, whose vanishing marks the position of the apparent horizon. In this section, we relate the quantum expectation value of this operator to the value of $\;-\langle \Phi_q^2 \rangle$, which does not vanish and remains negative  even at the would-be horizon when $\;r\to 1\;$.
Consequentially,
the formation of an apparent horizon  is prevented by particle production.

While keeping in mind the classical expression as derived in Eq.~(\ref{expect}),
let us now evaluate the quantum expectation value of the relevant product ({\em cf}, Eq.~(\ref{preexpect})),
$$
  -\langle\gamma^{\alpha\beta}\nabla_\alpha \Phi(\text{\textsc{u}},\text{\textsc{v}}) \nabla_\beta \Phi(\text{\textsc{u}},\text{\textsc{v}})\rangle\;=\;  r  e^{r} \langle\partial_\text{\textsc{u}}\Phi(\text{\textsc{u}},\text{\textsc{v}}) \partial_\text{\textsc{v}} \Phi(\text{\textsc{u}},\text{\textsc{v}})\rangle\;,
$$
where  $\;\Phi\Phi=\Phi_0\Phi_0+ \Phi_q\Phi_q \;$
is a sum over the square of the zero mode  and the square of each of the $\;\ell>0\;$ modes.

To evaluate this, it is convenient  to first  convert the $U$ and $V$ derivatives via
$\;\partial_\text{\textsc{v}}=\frac{1}{\text{\textsc{v}}}( \partial_{r^*}+\partial_t)\;$ and
$\;\partial_\text{\textsc{u}}=\frac{1}{\text{\textsc{u}}}( \partial_{r^*}-\partial_t)\;$,
so that
\bea
 && \langle\partial_\text{\textsc{u}}\Phi(\text{\textsc{u}},\text{\textsc{v}}) \partial_\text{\textsc{v}} \Phi(\text{\textsc{u}},V)\rangle=\frac{1}{\text{\textsc{u}}\text{\textsc{v}}}
 \langle \left( -(\partial_t\Phi)^2+(\partial_r^*\Phi)^2\right)\rangle
  \cr &=&
  -\frac{1}{(r-1) e^r}  \langle\left( -(\partial_t\Phi)^2+(\partial_r^*\Phi)^2\right)\rangle
  \cr &=&
  -\frac{1}{(r-1) e^r}\left( \omega^2-k^2\right) \langle\Phi^2\rangle
  \;,
\eea
where the second equality used Eq.~(\ref{uvinr}).

Recalling that $\;\omega^2=k^2+m^2\;$ and  $\;m^2(r^*)= \frac{r-1}{r^4}\;$ (see Eq.~(\ref{m2Rstar})), we find that
\bea
 && -\langle\gamma^{\alpha\beta}\nabla_\alpha \Phi(\text{\textsc{u}},\text{\textsc{v}}) \nabla_\beta \Phi(\text{\textsc{u}},\text{\textsc{v}})\rangle
\cr  &=& re^r\left[\langle\partial_U\Phi_0(\text{\textsc{u}},\text{\textsc{v}}) \partial_\text{\textsc{v}} \Phi_0(\text{\textsc{u}},\text{\textsc{v}})\rangle \right.
+ \left.\langle\partial_\text{\textsc{u}}\Phi_q(\text{\textsc{u}},\text{\textsc{v}}) \partial_\text{\textsc{v}} \Phi_q(\text{\textsc{u}},\text{\textsc{v}})\rangle\right]
\cr  &=& -\frac{r-1}{r}-\frac{1}{r^3}\langle\Phi_q^2\rangle_{4D}\;.
\eea
Notice that, away from the would-be horizon, the quantum contribution is suppressed so that the two terms become comparable only near $\;r=2MG\;$.

Using  Eq.~(\ref{varHHH}) and taking the would-be horizon  limit, we have
(with units restored and the original, non-canonical fields assumed)
\bea
 && -\langle\gamma^{\alpha\beta}\nabla_\alpha \Phi(\text{\textsc{u}},\text{\textsc{v}}) \nabla_\beta \Phi(\text{\textsc{u}},\text{\textsc{v}})\rangle_{\bigl|r=1}
 = -\langle\Phi_q^2\rangle_{4D\bigl| r=1}
 \cr
 &=&-\frac{(2MG)^2}{3}\ln{\left(\frac{r_i}{2MG}\right)}\;,
\eea
which is manifestly less than zero because $\;r_i\gg 2MG\;$.

The conclusion is that
$\langle\sqrt{-g}~\Theta_U \Theta_V\rangle$ remains negative and has a magnitude of at least order one times $(2MG)^2$ at $\;r=2MG\;$. Consequently, it is clear that a trapped surface never does form.

\section{Discussion}

What we have found is that  the quantum width
of the would-be horizon scales as the shell's gravitational radius $2MG$. This  result implies the blurring of the horizon over a significant scale, which obstructs the formation of an apparent horizon and a trapped surface.

One can deduce that, as the shell approaches its gravitational radius, the production of quanta for a given scalar mode is a perturbatively small effect, scaling as $l_P^2$.  However, as mentioned above, there is an exceptionally large number of accessible modes; so much so that the total number of produced quanta scales as the Bekenstein--Hawking entropy \cite{Bek,Haw}  and their energy scales as the mass of the shell itself \cite{shellgame}.

Our conclusion is then that the classical framework leading to the proof of  trapped-surface formation is invalidated.

Like Christodoulou and others, we have focused on the simple case of a spherically symmetric, thin shell of classical matter following along a null trajectory. Generalizations to more realistic matter systems are possible, but our expectation is that the basic findings  would still be realized. It would be of  interest to investigate the case of  non-zero angular momentum, to  which our methods could be readily applied.

An intriguing question which remains open  is the fate of the collapsing matter system. We have shown that the matter does not form a trapped surface so that the singularity theorems are invalidated, but then what object does form?

Given that a trapped surface does not form, one has to confront the issue that it is no longer possible to describe the interior of the would-be horizon as a region of  empty space as assumed when analytically continuing the classical Schwarzschild geometry. Since the backreaction is not strong enough to halt the collapse of the shell, the matter continues to contract with a mass that is approximately $M$ but with a radius that is  less than $2MG$. It is not possible in this case to connect the interior geometry to the classical Schwarzschild geometry;
meaning that the classical evolution equations, which rely on using the classical geometry, can no longer be applied to learn about the fate of the shell. The conclusion is that some extreme deviation  away from the classical geometry has to take place as the shell advances past $\;r=2MG\;$.

Our preferred resolution (see \cite{tun} and many of the references within) is that the result of the collapse is the formation of an ultracompact object, whose outer surface is parametrically close to the would-be horizon such that the exterior geometry has a very large but finite redshift. The object cannot further collapse and has to posses an entropy which is approximately the Bekenstein--Hawking value. Standard model matter does not possess these properties, so that the original substance has to transform, quantum mechanically, to some exotic form.

However, the identification of the specific dynamical mechanism which facilitates such a quantum phase transition on arbitrarily large scales remains the crux of the challenge.

\section*{Acknowledgments}

We thank Paulo Pani for forcing us to think harder about the problem of collapse, Sunny Itzhaki for sharing his insights about the fate of collapsing matter and for many discussions,  Mihalis Dafermos for helping us understand the framework of the proof of the trapped surface formation theorem and to Paul Steinhardt for highlighting the connection of our analysis to cosmology. We also thank the participants of the Simons Center for Geometry and Physics program ``50 years of the blackhole information paradox'' for comments and questions. The research of AJMM received support from an NRF Evaluation and Rating Grant 119411 and a Rhodes  Discretionary Grant SD07/2022. The research of RB and HM is supported by the German Research Foundation through a German-Israeli Project Cooperation (DIP) grant ``Holography and the Swampland'' and by VATAT (Israel planning and budgeting committee) grant for supporting theoretical high energy physics. RB thanks the theory department at CERN, the department of theoretical physics at the University of Geneva and the Arnold Sommerfeld Center of Ludwig Maximilians University, Munich,  for their hospitality. AJMM thanks Ben Gurion University for their hospitality during previous visits.

\appendix

\section{Crucial on Kruskal}

In general, one can express  a Kruskal-like null coordinate system
as
\begin{equation}
    ds^2 \;= \;2\gamma_{UV}dUdV +r^2 d\Omega^2\;,
\end{equation}
where $U\;\text{and } V$ are defined such that $\gamma_{UV}$ does not turn to zero or infinity over  any  finite period of  time.

For 2D Schwarzschild  in particular,
\begin{equation}
    ds^2 \;=\; -2 h dU dV \;=\; -f dt^2 +\frac{1}{f}dr^2 \;,
\end{equation}
where $\;f=1-\frac{1}{r}\;$ and, by convention, $U$ and $V$ are dimensionless whereas $h$ is dimensional. One finds that~~\footnote{To obtain these results, it is convenient to
start out with the relation between Kruskal and Schwarzschild tortoise coordinates, $\;V=e^{\frac{r^* +t}{2}}\;$
and  $\;U=-e^{\frac{r^* -t}{2}}\;$.
\label{wow}}
\begin{equation}
        r \;=\;  \left(1+PL\left(\frac{-UV}{e} \right) \right)\;,
        \label{rinuv}
\end{equation}
where $PL$ denotes the product logarithm function (also known as the
Lambert W function), meaning that $\;\frac{-UV}{e}=(r-1)e^{r-1}\;$ or
\begin{equation}
        UV \;=\; \left(1-r\right)e^{r}\;.
        \label{uvinr}
\end{equation}

The previous leads to
\begin{equation}
    ds^2 \;=\; -2 h dUdV \;=\; -\frac{4}{r} e^{-r} dUdV\;,
\end{equation}
that is,
\begin{equation}
    -\gamma_{UV}\;=\;h\; =\; \frac{2(2MG)^3}{r} e^{-r/2MG} \;.
\end{equation}

Alternatively, one can define the Kruskal temporal  and ``radial'' coordinates as $\;V=T+X\;$ and $\;U=T-X\;$ to obtain
\begin{equation}
ds^2 \;=\; \frac{4}{r}e^{-r} \left(-dT^2+dX^2\right)\;.
\end{equation}

Returning to Kruskal null coordinates, let us consider
the expansions of a pair of  null congruences, ingoing and outgoing,
respectively.
The expansion parameter $\Theta$ measures the rate at which the cross-sectional area of a bundle of geodesics changes. For a congruence that is characterized by a tangent vector field $K^\alpha$, the parameter can be expressed as
\begin{equation}
    \Theta = \frac{1}{\sqrt{-g}} \partial_\alpha \left( \sqrt{-g} \, K^{\alpha} \right),
\end{equation}
where $g$  is the determinant of the full spacetime metric. This scalar $\Theta$ plays an important role in the analysis of gravitational collapse.

If there is no angular momentum,
the  tangent vectors for each congruence of interest can be deduced from
\begin{equation}
    K_U \;=\; -\partial_\alpha V\;,   \;\;\;\;\; K_V\;=\;-\partial_\alpha U\;.
\end{equation}
 Then, since
\begin{equation}
    {K_U}_\alpha \;=\; (0,-1,0,0)\;, \; \; \; \; {K_V}_\alpha \;=\; (-1,0,0,0)\;,
\end{equation}
the corresponding tangent vectors are
\begin{equation}
    {K_U}^{\alpha} \;=\; (-\gamma^{UV},0,0,0)\;,  \; \; \; \; {K_V}^{\alpha} \;= \;(0,-\gamma^{UV},0,0)\;.
\end{equation}

Let us next consider that the square root of the four-dimensional  determinant goes as  (with $r$ replaced by  the scalar field $\Phi$)
$\;\sqrt{-g} = -\gamma_{UV}\Phi^2\;$. It then follows that
\bea
    \Theta_U &=& \frac{1}{\gamma_{UV}\Phi^2} \partial_\alpha \left(\gamma_{UV}\Phi^2 {K_U}^{\alpha}\right) \\ \nonumber & = & \frac{1}{\gamma_{UV}\Phi^2} \partial_U \left(-\gamma_{UV}\Phi^2 \gamma^{UV}\right)\nonumber \\ &=& -\frac{2}{\gamma_{UV}\Phi}\partial_U\Phi\;,
\eea
\bea
    \Theta_V &=& \frac{1}{\gamma_{UV}\Phi^2} \partial_\alpha \left(\gamma_{UV}\Phi^2 {K_V}^{\alpha}\right) \nonumber \\ & = & \frac{1}{\gamma_{UV}\Phi^2} \partial_V \left(-\gamma_{UV}\Phi^2 \gamma^{UV}\right) \nonumber \\ &=& -\frac{2}{\gamma_{UV}\Phi}\partial_V\Phi\;,
\eea
which yields the relation
\begin{equation}
\frac{1}{2}~\sqrt{-g}~\Theta_U \Theta_V\; =\;  -\gamma^{\alpha\beta}\nabla_\alpha r(U,V) \nabla_\beta r(U,V)\;.
\label{preexpect}
\end{equation}

The above product  must vanish
at an apparent horizon -- where, classically,  the redshift becomes infinite and the light cone degenerates
 --- being that the outer boundary
of a trapped region is a marginally trapped surface.
In the case of a Schwarzschild geometry, this is abundantly clear, as one finds that
\be
-\gamma^{\alpha\beta}\nabla_\alpha r(U,V) \nabla_\beta r(U,V)\;=\; -\frac{r-1}{r}\;,
\label{expect}
\ee
which indeed vanishes at $\;r=1\;$ and is negative for $\;r>1\;$ in the exterior.
However, as documented in the main text, we find that this never happens when
the angular modes are taken into account.

Finally, one should take note of
\begin{equation}
\square \Phi= \frac{1}{\sqrt{-\gamma}}\partial_\alpha\left(\sqrt{-\gamma} \gamma^{\alpha\beta} \partial_\beta \Phi \right)
   \; =\; -\frac{2}{h} \partial_U \partial_V \Phi\;.
\end{equation}


\begin{thebibliography}{99}


\bibitem{info}
S. W. Hawking,
``Breakdown of predictability in gravitational collapse,''
Phys. Rev. D {\bf 14}, 2460  (1976).







\bibitem{mathurCP}
S.~D.~Mathur,
``Resolving the black hole causality paradox,''
Gen. Rel. Grav. \textbf{51}, no.2, 24 (2019)
[arXiv:1703.03042 [hep-th]].




\bibitem{AMPS}
 A.~Almheiri, D.~Marolf, J.~Polchinski and J.~Sully,
  ``Black Holes: Complementarity or Firewalls?,''
  JHEP {\bf 1302}, 062 (2013)
  [arXiv:1207.3123 [hep-th]].



 \bibitem{BD}
 N.~D.~Birrell and  P.~C.~W.~Davies,
 {\em Quantum Fields in Curved Space}
 (Cambridge University Press, 1982).




\bibitem{Haw}
S. W. Hawking,
``Black hole explosions,'' Nature {\bf 248}, 30 (1974);
``Particle creation by black holes,'' Comm. Math. Phys. {\bf 43}, 199 (1975).




\bibitem{carded}
V.~Cardoso and P.~Pani,
``Testing the nature of dark compact objects: a status report,''
Living Rev. Rel. \textbf{22}, no.1, 4 (2019)
[arXiv:1904.05363 [gr-qc]].



\bibitem{thumper}
C.~Bambi, R.~Brustein, V.~Cardoso, A.~Chael, U.~Danielsson, S.~Giri, A.~Gupta, P.~Heidmann, L.~Lehner and S.~Liebling, \textit{et al.}
``Black hole mimickers: from theory to observation,''
arXiv:2505.09014 [gr-qc].




\bibitem{PenHawk1}
R. Penrose, ``Gravitational Collapse and Space-Time Singularities,''
Phys. Rev. Lett. {\bf 14}, 57 (1965).

\bibitem{PenHawk2}
S. W. Hawking and R. Penrose,
``The singularities of gravitational collapse and cosmology,''
Proc. R. Soc. Lond. A  {\bf 314},  529 (1970).




\bibitem{HawkEl}
S.~W.~Hawking and G.~F.~R. Ellis,
{\em The Large Scale Structure of Space-Time}
(Cambridge University Press, 1973).

\bibitem{sing}
J.~M.~M.~Senovilla,
``Singularity Theorems and Their Consequences,''
Gen. Rel. Grav. \textbf{30}, 701 (1998)
[arXiv:1801.04912 [gr-qc]].



\bibitem{HawkingEllis}
S.~W.~Hawking and G.~F.~R.~Ellis,
``The Large Scale Structure of Space-Time,''
Cambridge University Press.

\bibitem{OS}
J.~R. Oppenheimer and H. Snyder, ``On Continued Gravitational Contraction,''
Phys. Rev. {\bf 56}, 455-459 (1939).


\bibitem{MTW}
C.~W.  Misner, K.~S. Thorne and J.~A. Wheeler,
{\em Gravitation} (W.H. Freeman and Company, San Fransisco, 1973).




 \bibitem{Chris1}
D.~Christodoulou,
``The Problem of a Selfgravitating Scalar Field,''
Commun. Math. Phys. \textbf{105}, 337-361 (1986).

\bibitem{Chris2}
D.~Christodoulou,
``The formation of black holes and singularities in spherically symmetric gravitational collapse,''
Commun. Pure Appl. Math. \textbf{44}, no.3, 339-373 (1991).


\bibitem{Chris3}
D. Christodoulou,
``Bounded variation solutions of the spherically symmetric einstein-scalar field equations,''
Commun. Pure Appl. Math. {\bf 46} 1131-1220 (1993).

\bibitem{Chris4}
D.~Christodoulou,
``Examples of naked singularity formation in the gravitational collapse of a scalar field,''
Annals Math. \textbf{140}, 607-653 (1994).

\bibitem{Chris5}
D.~Christodoulou,
``The Instability of Naked Singularities in the Gravitational Collapse of a Scalar Field,''
Annals Math. \textbf{149}, no.1, 183 (1999)
[arXiv:math/9901147 [math.AP]].

\bibitem{Chris6}
 D.~Christodoulou,
 ``On the global initial value problem and the issue of singularities,''
 Class. Quantum Grav. {\bf 16} A23 (1999).

 \bibitem{Chris7}
D.~Christodoulou,
``The Formation of Black Holes in General Relativity,''
arXiv:0805.3880 [gr-qc].




\bibitem{numeric1}
M.~W.~Choptuik,
``Critical behaviour in massless scalar field collapse,''
in  {\em Approaches to Numerical Relativity}, R.~D’Inverno, ed.
(Cambridge University Press, 1992) pp. 202-222.

\bibitem{numeric2}
M.~W.~Choptuik,
``Universality and scaling in gravitational collapse of a massless scalar field,''
Phys. Rev. Lett. \textbf{70}, 9-12 (1993).


\bibitem{numrev}
C.~Gundlach and J.~M.~Martin-Garcia,
``Critical phenomena in gravitational collapse,''
Living Rev. Rel. \textbf{10}, 5 (2007)
[arXiv:0711.4620 [gr-qc]].


\bibitem{dust1}
T.~P.~Singh and P.~S.~Joshi,
``The Final fate of spherical inhomogeneous dust collapse,''
Class. Quant. Grav. \textbf{13}, 559-572 (1996)
[arXiv:gr-qc/9409062 [gr-qc]].

\bibitem{dust2}
F.~C.~Mena and B.~C.~Nolan,
``Nonradial null geodesics in spherical dust collapse,''
Class. Quant. Grav. \textbf{18}, 4531-4548 (2001)
[arXiv:gr-qc/0108008 [gr-qc]].




\bibitem{aniso1}
R.~Giambo, F.~Giannoni, G.~Magli and P.~Piccione,
``New solutions of Einstein equations in spherical symmetry: The Cosmic censor to the court,''
Commun. Math. Phys. \textbf{235}, 545-563 (2003)
[arXiv:gr-qc/0204030 [gr-qc]].


\bibitem{aniso2}
R.~Giambo, F.~Giannoni, G.~Magli and P.~Piccione,
``New mathematical framework for spherical gravitational collapse,''
Class. Quant. Grav. \textbf{20}, L75 (2003)
[arXiv:gr-qc/0212082 [gr-qc]].


\bibitem{charge1}
 M.~Dafermos,
``Stability and instability of the Cauchy horizon for the spherically symmetric Einstein-Maxwell-scalar field equations,''
Ann. Math. {\bf 158}, no.3, 875-928 (2003).


\bibitem{charge2}
M.~Dafermos,
``The interior of charged black holes and the problem of uniqueness in general relativity,''
Commun. Pure Appl. Math. \textbf{58}, no.4, 0445-0504 (2005)
[arXiv:gr-qc/0307013 [gr-qc]].



\bibitem{adSV}
T.~Hertog, G.~T.~Horowitz and K.~Maeda,
``Generic cosmic censorship violation in anti-de Sitter space,''
Phys. Rev. Lett. \textbf{92}, 131101 (2004)
[arXiv:gr-qc/0307102 [gr-qc]].





\bibitem{charge3}
M.~Dafermos and I.~Rodnianski,
``A Proof of Price's law for the collapse of a selfgravitating scalar field,''
Invent. Math. \textbf{162}, 381-457 (2005)
[arXiv:gr-qc/0309115 [gr-qc]].





\bibitem{higgsV}
M.~Dafermos,
``A Note on naked singularities and the collapse of selfgravitating Higgs fields,''
Adv. Theor. Math. Phys. \textbf{9}, no.4, 575-591 (2005)
[arXiv:gr-qc/0403033 [gr-qc]].





\bibitem{refine1}
J.~Luk and S.~J.~Oh,
``Quantitative decay rates for dispersive solutions to the Einstein-scalar field system in spherical symmetry,''
Anal. Part. Diff. Eq. \textbf{8}, no.7, 1603-1674 (2015)
[arXiv:1402.2984 [gr-qc]].


\bibitem{refine2}
J.~Luk, S.~J.~Oh and S.~Yang,
``Solutions to the Einstein-scalar-field system in spherical symmetry with large bounded variation norms,''
arXiv:1605.03893 [gr-qc].


\bibitem{refine3}
J.~Liu and J.~Li,
``A robust proof of the instability of naked singularities of a scalar field in spherical symmetry,''
Commun. Math. Phys. \textbf{363}, no.2, 561-578 (2018)
[arXiv:1710.02922 [gr-qc]]






\bibitem{dSV}
J.~L.~Costa,
``The formation of trapped surfaces in the gravitational collapse of spherically symmetric scalar fields with a positive cosmological constant,''
Class. Quant. Grav. \textbf{37}, no.19, 195022 (2020)
[arXiv:2005.03434 [gr-qc]].















\bibitem{review}
R.~Giamb{\`o},
``Gravitational Collapse of~a~Spherical Scalar Field,''
in {\em New Frontiers in Gravitational Collapse and Spacetime Singularities},
 P. Joshi and D. Malafarina, eds.
 (Springer, Singapore, 2024) pp. 141-173
[arXiv:2307.01796 [gr-qc]].




\bibitem{hop}
N.~Itzhaki,
``The Horizon order parameter,''
[arXiv:hep-th/0403153 [hep-th]].




\bibitem{paddy}
A.~Paranjape and T.~Padmanabhan,
``Radiation from collapsing shells, semiclassical backreaction and black hole formation,''
Phys. Rev. D \textbf{80}, 044011 (2009)
[arXiv:0906.1768 [gr-qc]].


\bibitem{manbear}
P.~Chen, W.~G.~Unruh, C.~H.~Wu and D.~H.~Yeom,
``Pre-Hawking radiation cannot prevent the formation of apparent horizon,''
Phys. Rev. D \textbf{97}, no.6, 064045 (2018)
[arXiv:1710.01533 [gr-qc]].

\bibitem{shellgame}
R. Brustein, A.J.M. Medved and H. Meir,
``Suppression of trapped surface formation by quantum gravitational effects,''
longer version of current letter, to be submitted in tandem.



\bibitem{reduction1}
Y.~M.~Cho, K.~S.~Soh, J.~H.~Yoon and Q.~H.~Park,
``Gravitation as gauge theory of diffeomorphism group,''
Phys. Lett. B \textbf{286}, 251-255 (1992)

\bibitem{reduction2}
J.~H.~Yoon,
``Four-dimensional Kaluza-Klein approach to general relativity in the (2,2) splitting of space-times,''
[arXiv:gr-qc/9611050 [gr-qc]].

\bibitem{reduction3}
J.~H.~Yoon,
``Kaluza-Klein formalism of general space-times,''
Phys. Lett. B \textbf{451}, 296-302 (1999)
[arXiv:gr-qc/0003059 [gr-qc]].

\bibitem{mycft}
R.~Brustein,
``Causal boundary entropy from horizon conformal field theory,''
Phys. Rev. Lett. \textbf{86}, 576-579 (2001)
[arXiv:hep-th/0005266 [hep-th]].

\bibitem{carlip}
S.~Carlip,
``Near-horizon Bondi-Metzner-Sachs symmetry, dimensional reduction, and black hole entropy,''
Phys. Rev. D \textbf{101}, no.4, 046002 (2020)
[arXiv:1910.01762 [hep-th]].


 \bibitem{offord}
 C.~Pathinayake and L.~H.~Ford,
``Particle creation by a self-coupled scalar field,''
Phys. Rev. D {\bf 35}, 3709 (1987).




\bibitem{Bek}
  J.~D.~Bekenstein,
  ``Black holes and entropy,''
  Phys.\ Rev.\ D {\bf 7}, 2333 (1973).


\bibitem{tun}
R.~Brustein, A.~J.~M.~Medved and T.~Simhon,
``Formation of Frozen Stars from collapsing matter by tunneling,''
[arXiv:2508.02100 [gr-qc]].





\end{thebibliography}
\end{document}